\documentclass[showpacs,preprintnumbers,amsmath,amssymb]{revtex4}

\usepackage{graphicx}% Include figure files
\usepackage{dcolumn}% Align table columns on decimal point
\usepackage{bm}% bold math

\def\simge{\mathrel{%
       \rlap{\raise 0.511ex \hbox{$>$}}{\lower 0.511ex \hbox{$\sim$}}}}
\def\simle{\mathrel{
       \rlap{\raise 0.511ex \hbox{$<$}}{\lower 0.511ex \hbox{$\sim$}}}}

\begin{document}

\preprint{\sf BNL-NT-07/29\ \ 2007/June}

\title{On the existence of the critical point 
in finite density lattice QCD}

\author{Shinji Ejiri}

\affiliation{Physics Department, Brookhaven National Laboratory,
Upton, New York 11973, USA}

%\date{\today}
%\date{June 24, 2007}
\date{December 6, 2007}

\begin{abstract}
We propose a method to probe the nature of phase transitions in lattice 
QCD at finite temperature and density, which is based on the investigation 
of an effective potential as a function of the average plaquette. 
We analyze data obtained in a simulation of two-flavor QCD using 
p4-improved staggered quarks with bare quark mass $m/T = 0.4$, 
and find that a first order phase transition line 
appears in the high density regime for $\mu_q/T \simge 2.5$. 
We also discuss the difference between the phase structures of QCD 
with non-zero quark chemical potential and non-zero 
isospin chemical potential.
\end{abstract}


\maketitle

\section{Introduction}
\label{sec:intro}

In the last several years remarkable progress has been made in numerical 
studies of lattice QCD at finite temperature $(T)$ and quark chemical 
potential $(\mu_q)$.
The transition line, separating hadron phase and quark-gluon 
plasma (QGP) phase, was investigated from $\mu_q=0$ to finite $\mu_q$ 
\cite{FK1,FK2,BS02,dFP1,dEL1}, 
and the equation of state was also computed 
at low density \cite{BS02,BS03,BS05,isen06,MI06}. 
Among others, the study of the endpoint of the first order phase 
transition line in the $(T, \mu_q)$ plane is particularly important 
both from the experimental and theoretical point of view. 
This existence of such a critical point 
is suggested by phenomenological studies \cite{AY,Bard,SRS}.
The appearance of the critical endpoint in the $(T, \mu_q)$ plane 
is closely related to hadronic fluctuations in heavy ion collisions 
and may be experimentally examined by an event-by-event analysis of 
heavy ion collisions.

Although many trials have been made to prove the existence of 
the critical endpoint by first principle calculation in 
lattice QCD, no definite conclusion on this issue is obtained so far.
The first trial to find the critical endpoint by numerical 
simulations was performed in Ref.~\cite{FK2} investigating 
the finite size scaling behavior of Lee-Yang zeros 
in the complex $\beta=6/g^2$ plane. 
The difficulty in the Lee-Yang zero method for finite density QCD 
is discussed in Ref.~\cite{ej06}.
The radius of convergence in the framework of the Taylor expansion 
of the grand canonical potential can establish a lower bound on 
the location of the critical endpoint \cite{BS03,BS05,GG2}.
There are also studies in which the behavior of the critical endpoint 
as a function of the quark masses is examined 
by using the property that a critical endpoint exists at 
$\mu_q=0$ in the very small quark mass region for QCD with three 
flavors having degenerate quark masses \cite{crtpt,dFP2,dEL2}.
Moreover, studies by simulations of phase-quenched finite density QCD, 
have been performed in Ref.~\cite{KS05,KS06,FKS07}.

The purpose of this study is to clarify the existence of 
the endpoint of the first order phase transition line 
in the $(T, \mu_q)$ plane. 
We propose a new method to investigate the nature of transition.
In the study of finite density lattice QCD, 
the reweighting method \cite{Swen88,Bar} plays an important role. 
However, the calculation of physical quantities becomes 
increasingly more difficult for large $\mu_q$ 
due to the sign problem \cite{spli,ej04}. 
We also consider a way to avoid the sign problem. 
%We propose an approximation, which is valid when the volume 
%is sufficiently large, to be free from the sign problem. 

We evaluate an effective potential as a function of 
the average plaquette, and identify the type of 
transition from the shape of the potential. 
The partition function can be written as 
\begin{eqnarray}
{\cal Z}(\beta, \mu_q) = \int R(P,\mu_q) w(P) e^{-S_g(P,\beta)} \ dP,
\label{eq:rewmuP}
\end{eqnarray}
where $P$ denotes the plaquette value, $S_g(P,\beta)$ is the gauge action, 
$w(P)$ is the state density at $\mu_q=0$ for each $P$, 
and $R(P,\mu_q)$ is the modification factor for finite $\mu_q$.
$R(P,\mu_q)$ is obtained by calculating the quark determinant $\det M$ 
and is assumed to be real and positive. 
We then define the effective potential as 
$V(P, \beta, \mu_q) =- \ln (R w e^{-S_g})$. 
If there is a first order phase transition point, where two different 
states coexist, the potential must have two 
minima at two different values of $P$. 
However, the calculation of the quark determinant is quite expensive 
and is actually difficult except on small lattices. Moreover, 
the sign problem is serious when we calculate $R(P,\mu_q)$ directly. 

This study is based on the following two ideas to avoid these problems. 
One is that we perform a Taylor expansion of $\ln \det M(\mu_q)$ 
in terms of $\mu_q$ around $\mu_q=0$ and calculate the expansion coefficients, 
as proposed in Ref.~\cite{BS02}.
The Taylor expansion coefficients are rather easy to calculate 
by using the stochastic noise method. 
Although we must cut off this expansion at an appropriate order 
in $\mu_q$, we can estimate the application range where 
the approximation is valid for each analysis. 
While the application range of the Taylor expansion of $\ln {\cal Z}$ 
should be limited by the critical point because $\ln {\cal Z}$ 
is singular at the critical point, there is no such limit for 
the application range in the expansion of $\ln R(P,\mu_q)$ 
because the weight factor should always be well-defined. 
This discussion is given in Sec.~\ref{sec:taylor}.

The second idea is that 
we consider the probability distribution function 
in terms of the complex phase of the quark determinant 
$\theta$ when $P$ and $|\det M|$ are fixed.
We assume the distribution function is well-approximated 
by a Gaussian function, and perform the integration 
over $\theta$. If we adopt this assumption, the sign problem 
in the calculation of $\ln R(P,\mu_q)$ is completely solved. 
This assumption is reasonable for sufficiently large volume 
and is suggested by the simulation results given 
in this study. We discuss this method in Sec.~\ref{sec:sign}. 

General remarks on the phase transition in lattice QCD are 
given in Sec.~\ref{sec:tran}, and an effective potential as a 
function of the average plaquette is introduced. 
We discuss the reweighting method for the study of 
the QCD phase structure at non-zero temperature and density 
in Sec.~\ref{sec:rew}. We evaluate the effective potential 
using data obtained with two-flavors of p4-improved 
staggered quarks in Ref.~\cite{BS05}. 
We also discuss the phase structure of QCD with isospin chemical potential.
Our conclusions are given in Sec.~\ref{sec:conc}.

\section{Probability distribution function and phase transition}
\label{sec:tran}

%\paragraph*{Grand canonical partition function}
The grand canonical partition function of lattice QCD is given by
\begin{eqnarray}
{\cal Z} (\beta, \mu_q) = \int 
{\cal D}U \left( \det M \right)^{N_{\rm f}} e^{-S_g},
\label{eq:partition} 
\end{eqnarray}
and the expectation value of an operator ${\cal O}$ is calculated by 
\begin{eqnarray}
\langle {\cal O} \rangle 
= \frac{1}{\cal Z} \int {\cal D}U 
{\cal O} (\det M)^{N_{\rm f}} e^{-S_g},
\label{eq:expectation} 
\end{eqnarray}
where $M(\mu_q)$ is the quark matrix. $N_{\rm f}$ is the number of flavors. 
When one use a staggered type quark action, $N_{\rm f}$ is 
replaced by $ N_{\rm f}/4$. $S_g(\beta)$ is the gauge action, which is given 
by a linear combination of the Wilson loops 
$W^{I \times J}_{\mu \nu} (x)$, where $I \times J$, $\mu \nu$ and $x$ 
are the size, direction and position of the Wilson loop, respectively, 
$\beta$ is a simulation parameter related to the gauge coupling $g$ 
being $\beta=6/g^2$. 
The simplest gauge action is the standard plaquette action given 
by the following equation,
\begin{eqnarray}
S_g= -\beta \sum_{x, \mu > \nu} W^{1 \times 1}_{\mu \nu} (x).
\label{eq:sgplaq}
\end{eqnarray}
Because the $1 \times 1$ Wilson loop is defined on an elementary 
square (plaquette), $W^{1 \times 1}_{\mu \nu}$ is usually called 
plaquette or plaquette variable.

In a Monte Carlo simulation, we generate configurations of link variables 
$\{ U_{\mu}(x) \}$ with the probability in proportion to the weight factor 
$(\det M)^{N_{\rm f}} e^{-S_g}$ and the state density of $\{ U_{\mu}(x) \}$. 
The expectation value is then estimated by taking an average 
of the operator ${\cal O}[U_{\mu}]$ over the generated configurations 
$\{ U_{\mu}(x) \}$.
\begin{eqnarray}
\langle {\cal O} \rangle_{(\beta)} 
\approx \frac{1}{N_{\rm conf.}} \sum_{ \{ U_{\mu}(x) \} } {\cal O}[U_{\mu}].
\end{eqnarray}

%\paragraph*{Probability distribution function}
We introduce a probability distribution function 
of the plaquette, $w(P)$, which is defined by 
\begin{eqnarray}
w(P') = \int {\cal D} U \ \delta(P'-P) \ (\det M)^{N_{\rm f}} 
e^{6\beta N_{\rm site} P}, 
\label{eq:pdist}
\end{eqnarray}
where $\delta(x)$ is the delta function. 
For later discussions, we define the average plaquette $P$ as 
$ P \equiv -S_g/(6 \beta N_{\rm site})$.
This is the average of the plaquette over all elementary squares 
for the standard gauge action, Eq.~(\ref{eq:sgplaq}).
$N_{\rm site} = N_s^3 \times N_t$ is the number of sites.
Using the distribution function, 
the expectation value can be rewritten as 
\begin{eqnarray}
\langle {\cal O}[P] \rangle_{(\beta)} 
= \frac{1}{\cal Z} \int {\cal O}[P] \ w(P) \ d P, \hspace{5mm}
{\cal Z} = \int w(P) \ d P,  
\label{eq:expp}
\end{eqnarray}
for an operator given by the plaquette ${\cal O}[P]$. 
In the calculation of Eq.~(\ref{eq:pdist}), 
we actually use an approximate delta function 
such as a box type function, 
$\delta(x) \approx \{ 1/\Delta \ 
({\rm for} \ \Delta/2<x \leq \Delta/2), 
0 \ {\rm (otherwise)} \}$, 
or a Gaussian function,  
$\delta(x) \approx 1/(\Delta \sqrt{\pi}) \exp[-(x/\Delta )^2]$.
For the case of the box type, we can estimate $w(P)$ by counting 
the number of configurations for each value of $P$ with 
the width of box $\Delta$. As $\Delta$ decreases, the approximation 
becomes better but the statistical error becomes 
large because the number of configurations in each block 
becomes small. 
Hence, we must adjust the size of $\Delta$ appropriately.

Next, we discuss the shape of the probability distribution function.
In general, the number of states increases exponentially 
as the gauge fields become random. 
On the other hand, the random configurations are exponentially 
suppressed by the weight factor $\exp(6\beta N_{\rm site} P)$, 
since the plaquette is one when the gauge field is $U_{\mu}(x)=1$ 
uniformly (free gas limit) and $P$ decreases as the configuration 
becomes random.
Therefore, the most probable $P$ is determined by 
the balance of the number of states and the weight factor,
and the value of plaquette variable distributes around the most 
probable value for each point and each configuration. 

We first consider the case that there is no spatial correlation 
between the plaquette variables at each point and 
the volume is sufficiently large. 
In this case, the shape of the probability distribution 
as a function of the plaquette averaged over the space 
must be a Gaussian function. 
The central limit theorem tells us that the probability 
distribution of the average of the random numbers which have 
the same probability distribution is always of Gaussian type 
when the set of random numbers is large enough. 
We can apply this theorem in this case. Hence, 
\begin{eqnarray}
w(P)= \sqrt{\frac{6N_{\rm site}}{2\pi \chi_P}} 
\exp \left\{ -\frac{6N_{\rm site}}{2 \chi_P} 
\left( P- \langle P \rangle \right)^2 \right\}, 
\label{eq:gdis}
\end{eqnarray}
where $\langle P \rangle $ is the expectation value of $P$ and 
$\chi_P$ is the susceptibility, 
\begin{eqnarray}
\langle P \rangle = \int P \ w(P) \ dP, \hspace{5mm}
\chi_P \equiv 6N_{\rm site} \langle (P - \langle P \rangle)^2 \rangle
= 6N_{\rm site} \int (P - \langle P \rangle)^2 w(P) \ dP.
\label{eq:gcoe}
\end{eqnarray}

We expect that $w(P)$ is of Gaussian type also for more general 
interacting cases when the correlation 
length is much shorter than the size of the system. 
If we divide the space into domains which are larger than 
the correlation length and average the plaquette variables in 
these domains, the averaged plaquettes can be independent 
for each domain. When the number of domains is large, 
the distribution function as a function of the plaquette 
averaged over space must be a Gaussian function. 

However, we do expect that the probability distribution function is not 
of Gaussian type for the following two cases. 
One is, of course, the case that the correlation length is not small 
in comparison to the size of the system because the above-mentioned 
argument cannot be applied. 
The other case is that the most probable values of plaquette 
is not unique. For this case, the whole space is separated 
into domains having different states, 
and the plaquette variables in each domain distribute around one of 
the most probable values of plaquette. 
Although, on the surface separating these domains, the most probable 
plaquette value may not be realized, the effect from the wall becomes 
smaller as the volume increases, since the effect from the wall 
increases as a function of the area of the wall. 
Consequently, the existence of the domain wall does 
not affect the probability in the infinite volume limit. 
The probability distribution function should then be flat 
in the range between these most probable values of $P$ because 
the spatial average of $P$ depends on the size of these domains but 
the probability does not change in this range. 
However, in a finite volume, the effect from the domain wall 
cannot be neglected, hence the distribution function has two peaks 
when the number of most probable values for $P$ is two. 
Clearly the two exceptions discussed here correspond 
to the case at a second order phase transition point 
and at a first order phase transition point, respectively.

%\paragraph*{effective potential} 
Here, it is convenient to introduce the effective potential defined by
\begin{eqnarray}
V(P) = -\ln w(P).
\end{eqnarray}
As discussed above, the distribution function is normally written as 
$w(P) \sim \exp \{ -(6N_{\rm site}/2 \chi_P) (P- \langle P \rangle)^2 \}$.
When one considers a Taylor expansion of $V(P)$ around the minimum 
$\langle P \rangle$, where the slope of the potential $dV/dP$ is zero, 
the effective potential is dominated by the second order 
term in the region near the minimum, i.e. 
the potential is a quadratic function in the vicinity of 
$\langle P \rangle$, and the second derivative (curvature) 
of $V(P)$ at $\langle P \rangle$ is related to 
the plaquette susceptibility with 
\begin{eqnarray}
\frac{d^2 V}{dP^2} = \frac{6N_{\rm site}}{\chi_P}.
\label{eq:curpot}
\end{eqnarray}

A second order phase transition point is characterized by 
the slope and curvature of the effective potential. 
The slope $dV/dP$ and curvature $d^2V/dP^2$ become zero 
simultaneously at the critical point.
As given in Eq.~(\ref{eq:gcoe}), $\chi_P$ is an indicator of 
fluctuations and diverges at a second order phase transition point
in the thermodynamic limit.
When the susceptibility $\chi_P$ becomes large in the vicinity of 
a second order phase transition point, the effect from 
the second order term of $V(P)$ becomes small 
in comparison to the higher order terms, and then 
the distribution function deviates from a Gaussian function. 
On the other hand, in the case of a first order phase transition point, 
more than one peak exist in the distribution function. This means that
there are points which give $dV/dP=0$ more than once, 
and the curvature of $V(P)$ may be negative 
around the mean value of $P$. 

%\paragraph*{Binder cumulant}
At the end of this section, we should also discuss the relation 
between the plaquette distribution function and 
the fourth order Binder cumulant, 
\begin{eqnarray}
B_4 \equiv 
\frac{\left\langle (P- \langle P \rangle)^4 \right\rangle}
{\left\langle (P- \langle P \rangle)^2 \right\rangle^2}, 
\label{eq:binder}
\end{eqnarray}
which often is used to identify the nature of a phase transition
\cite{KLS01}.
The value of the Binder cumulant at a second order critical point 
depends on the universality class. 
In the case of a first order phase transition, 
assuming the plaquette distribution is a double peaked function, 
the Binder cumulant is estimated as 
\begin{eqnarray}
B_4 = \frac{\int (P- \langle P \rangle)^4 w(P) \ dP}
{\left( \int (P- \langle P \rangle)^2 w(P) \ dP \right)^2}
\approx \frac{\Delta^4}{(\Delta^2)^2} \approx 1, 
\label{eq:b4dp}
\end{eqnarray}
where the distance between two peaks is $2\Delta$ and is wider 
than the width of each peak. 
On the other hand, when the distribution function can be modeled 
by a Gaussian function for a crossover transition or 
at a normal point, the Binder cumulant is given by 
\begin{eqnarray}
B_4 \approx \frac{\sqrt{x/\pi} \int (P- \langle P \rangle)^4 
e^{-x (P - \langle P \rangle )^2} dP}
{\left(\sqrt{x/\pi}\int (P- \langle P \rangle)^2 
e^{-x (P - \langle P \rangle )^2} dP \right)^2}
= \left(\sqrt{\frac{x}{\pi}} \frac{d^2 \sqrt{\pi/x}}{d x^2}
\right) \left/ \left(- \sqrt{\frac{x}{\pi}} 
\frac{d \sqrt{\pi/x}}{d x} \right)^2 \right.
= 3.
\label{eq:b4dp2}
\end{eqnarray}
In a region where a first order phase transition changes to 
a crossover, the Binder cumulant changes rapidly from one to three. 
We expect to find such a region for full QCD at high temperature and density. 

In addition, the method of Lee-Yang zeros has been used 
to identify the nature of the phase transition. 
The relation between the plaquette distribution function and 
the scaling analysis of the Lee-Yang zero 
has been discussed in Ref.~\cite{ej06}. The scaling behavior of 
the Lee-Yang zero can be also explained by 
the plaquette distribution function.
Hence, the distribution function of the plaquette plays an important 
role in the investigation of the nature of a phase transition.

\section{Lattice QCD at finite density}
\label{sec:rew}

The most difficult problem for lattice studies at finite baryon density is 
that the Boltzmann weight is complex when the chemical potential is non-zero. 
In this case, the Monte-Carlo method is not applicable directly, since 
configurations cannot be generated with a complex probability.
A popular approach to avoid this problem is the reweighting method. 
We perform simulations at $\mu_q=0$, and incorporate the remaining
part of the correct Boltzmann weight for finite $\mu_q$ in the calculation
of expectation values. 
Expectation values $\langle {\cal O} \rangle$ at $(\beta, \mu_q)$ are thus
computed by a simulation at $(\beta_0, 0)$ using the following identity, 
\begin{eqnarray}
\langle {\cal O} \rangle_{(\beta, \mu_q)} 
= \frac{\left\langle {\cal O} 
e^{N_{\rm f} (\ln \det M(\mu_q) - \ln \det M(0))}
e^{6 (\beta - \beta_0) N_{\rm site} P}
\right\rangle_{(\beta_0,0)}}{ \left\langle
e^{N_{\rm f} (\ln \det M(\mu_q) - \ln \det M(0))}
e^{6 (\beta - \beta_0) N_{\rm site} P}
\right\rangle_{(\beta_0,0)}}. 
\label{eq:murew}
\end{eqnarray} 
%where $M$ is the quark matrix and $N_{\rm f}$ is the number of flavors 
%($N_{\rm f}/4$ for staggered type quarks instead of $N_{\rm f}$); 
%$\mu$ is a quark chemical potential in lattice units, i.e. 
%$\mu \equiv \mu_q a=\mu_q/(N_{\tau} T)$, and $\mu_q$ is the quark 
%chemical potential in physical units.
This is the basic formula of the reweighting method. 
However, because $\ln \det M(\mu_q)$ is complex, 
the calculations of the numerator and denominator in Eq.~(\ref{eq:murew}) 
becomes in practice increasingly more difficult for larger $\mu_q$. 
We define the phase of the quark determinant $\theta$ by 
the imaginary part of $N_{\rm f} \ln \det M(\mu_q)$. 
%$(\det M(\mu_q))^{N_{\rm f}} \equiv |\det M(\mu_q)|^{N_{\rm f}} 
%e^{i \theta}$, 
If the typical value of $\theta$ becomes larger than $\pi/2$, the real 
part of $e^{i \theta}$ $(=\cos \theta)$ changes its sign frequently.
Eventually both the numerator and denominator of Eq.~(\ref{eq:murew}) 
become smaller than their statistical errors and Eq.~(\ref{eq:murew}) 
can no longer be evaluated. We call it the ``sign problem''.
The sign problem becomes more serious when the volume is large 
and the quark mass is small \cite{spli,ej04}.

\subsection{Reweighting method for finite $\mu_q/T$}

Let us discuss the reweighting method for finite $\mu_q$ 
using the plaquette distribution function. 
Originally, the reweighting method was proposed 
using the distribution function (histogram) in Ref.~\cite{Swen88}, 
and applications to the finite density QCD in this style 
have been discussed in Ref.~\cite{FKS07,Gock88,JN01,Ta04}. 

Here and hereafter, we restrict ourselves to discuss only the case when 
the quark matrix does not depend on $\beta$ explicitly, e.g. the standard 
Wilson and staggered quark actions, the p4-improved staggered 
quark action etc., for simplicity.
The partition function can be rewritten as 
\begin{eqnarray}
{\cal Z}(\beta, \mu_q) = \int R(P,\mu_q) w(P,\beta) \ dP,
\label{eq:rewmu}
\end{eqnarray}
where $w(P,\beta)$ is defined in Eq.~(\ref{eq:pdist}) at $\mu_q=0$ and 
$R(P,\mu_q)$ is the reweighting factor for finite $\mu_q$ 
defined by 
\begin{eqnarray}
R(P',\mu_q) \equiv 
\frac{\int {\cal D} U \ \delta(P'-P) (\det M(\mu_q))^{N_{\rm f}}}{
\int {\cal D} U \ \delta(P'-P) (\det M(0))^{N_{\rm f}}}.
\label{eq:rmudef}
\end{eqnarray}

This $R(P, \mu_q)$ is independent of $\beta$, and 
$R(P, \mu_q)$ can be measured at any $\beta$ 
using the following identity,
\begin{eqnarray}
R(P',\mu_q) 
= \frac{\int {\cal D} U \ \delta(P'-P) (\det M(\mu_q))^{N_{\rm f}} 
e^{6\beta N_{\rm site} P}}{
\int {\cal D} U \ \delta(P'-P) (\det M(0))^{N_{\rm f}} 
e^{6\beta N_{\rm site} P}} 
= \frac{ \left\langle \delta(P'-P) 
\left( \det M(\mu_q) / \det M(0) \right)^{N_{\rm f}}
\right\rangle_{(\beta, \mu_q=0)} }{
\left\langle \delta(P'-P) \right\rangle_{(\beta, \mu_q=0)}},
\label{eq:rmu}
\end{eqnarray}
where $\left\langle \cdots \right\rangle_{(\beta, \mu_q=0)}$ means 
the expectation value at $\mu_q=0$. In this method, 
all simulations are performed at $\mu_q=0$ and the effect of 
finite $\mu_q$ is introduced though the operator 
$\det M(\mu_q) / \det M(0)$ measured on the configurations 
generated by the simulations at $\mu_q=0$.
The expectation value of ${\cal O}[P]$ is given by 
\begin{eqnarray}
\langle {\cal O}[P] \rangle_{(\beta, \mu_q)} 
= \frac{\int {\cal O}[P] R(P, \mu_q) w(P,\beta) \ dP}
{\int R(P, \mu_q) w(P,\beta) dP}.
\end{eqnarray}
%If one calculates $R(P, \mu_q)$ precisely, 
%$\langle {\cal O}[P] \rangle$ can be evaluated at any $\mu_q$. 
Moreover, the weight factor for non-zero $\mu_q$ is 
$R(P, \mu_q) w(P,\beta)$, and thus the effective potential is defined by 
\begin{eqnarray}
V(P, \beta, \mu_q) \equiv -\ln [R(P, \mu_q) w(P, \beta)] 
= -\ln R(P, \mu_q) + V(P, \beta, 0).
\label{eq:potmu}
\end{eqnarray}
The shape of the effective potential can then also be investigated 
at non-zero $\mu_q$ once $R(P, \mu_q)$ is obtained.
%Because the curvature of $V(P, \beta, 0)$ at $\mu_q=0$ is positive 
%and the curvature of $V(P, \beta, \mu_q)$ at a second order phase 
%transition point is zero, 
%the parameter range where $-\ln R(P, \mu_q)$ has negative 
%curvature is required for the existence of the critical point.

However, there are two problems to calculate $R(P, \mu_q)$.
The first problem is that the calculation of the quark 
determinant $\det M(\mu_q)$ is very expensive. 
With present day computer resources, 
the exact calculation of $\det M(\mu_q)$ is difficult except 
on small lattices.
The second problem is the ``sign problem''. 
Because $\ln \det M(\mu_q)$ is complex, 
the calculations of the numerator of Eq.~(\ref{eq:rmu}) 
becomes in practice increasingly more difficult for larger $\mu_q$. 
If the complex phase factor of the quark determinant 
${\rm Re}[e^{i \theta}]$ changes its sign frequently, 
the expectation value of $R(P, \mu_q)$ becomes smaller than 
its statistical error and the calculation of 
$-\ln R(P, \mu_q)$ in the effective potential becomes impossible.

\subsection{Taylor expansion in terms of $\mu_q/T$} 
\label{sec:taylor}

To avoid the first problem, we perform a Taylor expansion 
in terms of $\mu_q$ around $\mu_q=0$ and calculate the expansion 
coefficients, as proposed in Ref.~\cite{BS02}.
We expand $\ln \det M(\mu_q)$ in a Taylor series, 
\begin{eqnarray}
\ln \left[ \frac{\det M(\mu_q)}{\det M(0)} \right] 
= \sum_{n=1}^{\infty} \frac{1}{n!} 
\left[ \frac{\partial^n (\ln \det M)}{\partial (\mu_q/T)^n} \right] 
\left( \frac{\mu_q}{T} \right)^n . 
\label{eq:detTay}
\end{eqnarray}
The Taylor expansion coefficients are rather easy to calculate 
by using the stochastic noise method. 
Although we must cut off this expansion at an appropriate order 
of $\mu_q$, this approximation is valid at low density and can be 
systematically improved by increasing the number of the terms.

Here, we discuss the effect of a truncation of the expansion.
To estimate the range of $\mu_q/T$ where the approximation is valid, 
an analysis of the radius of convergence is useful. 
The radius of convergence for pressure $p(T, \mu_q)$ is studied 
in Ref.~\cite{BS03}. When one performs a Taylor expansion for  
$p/T^4= \ln {\cal Z}/(VT^3)$, 
\begin{eqnarray}
\frac{p(T, \mu_q)}{T^4} - \frac{p(T, 0)}{T^4} 
=\sum_{i=1}^{\infty} c_i(T) \left( \frac{\mu_q}{T} \right)^{i}, 
\end{eqnarray}
the radius of convergence can be defined by 
\begin{eqnarray}
\rho= \lim_{i \to \infty} \rho_i, \hspace{5mm} 
\rho_{i}=\sqrt{ \left| \frac{c_{i}}{c_{i+2}} \right| }
\label{eq:conv}
\end{eqnarray}
for $i=2,4,6,\cdots$, where the odd terms are zero because 
the partition function is an even function of $\mu_q$.
The Taylor expansion converges in the range of $\mu_q/T < \rho$ 
when we consider all order of the expansion coefficients. 
This radius of convergence determine 
the lower bound of the critical point.
This means conversely that the upper limit of the application 
range must be below the critical point if we estimate thermodynamic 
quantities using the Taylor expansion coefficients of the pressure 
or $\ln {\cal Z}$. 

However, this problem may be avoidable 
when we consider a Taylor expansion of the reweighting factor 
$R(P, \mu_q)$ in Eq.~(\ref{eq:rewmu}), since the weight factor itself 
should not be singular even at the critical point. 
Therefore, we expect that the application range is not limited by 
the critical point and evaluations beyond the critical point is possible.
The same discussion of the radius of convergence is possible 
for the reweighting factor $\ln R(P, \mu_q)$. 
We define the expansion coefficients by
\begin{eqnarray}
\ln R(P, \mu_q) &=& 
\sum_{i=1}^{\infty} r_i (P) \left( \frac{\mu_q}{T} \right)^{i}, 
\end{eqnarray}
where the odd terms should be zero again. 
The radius of convergence is 
\begin{eqnarray}
\rho_{i}^{(R)} =\sqrt{ \left| \frac{r_{i}}{r_{i+2}} \right| }.
\label{eq:convrn}
\end{eqnarray}

When we neglect terms higher than $O(\mu_q^n)$ 
in the calculation of $\ln \det M$, Eq.~(\ref{eq:detTay}), 
the application range can be estimate by 
$\mu_q/T \simle \rho_n^{(R)}$. 
Because this approximation does not affect calculations 
of $r_i(P)$ for $i \le n$, the truncation error is negligible 
when the contribution from higher order terms is smaller than 
that from the lower order terms. 
In the range where $\mu_q/T < \rho_n^{(R)}$, the $(n+2)^{\rm th}$ 
order term $|r_{n+2} (\mu_q/T)^{n+2}|$
is smaller than the $n^{\rm th}$ order term $|r_n (\mu_q/T)^n|$. 
Hence, the truncation error must be small in this range.

Before discussing the radius of convergence for $\ln R(P,\mu_q)$ 
using the data obtained by Monte Carlo simulations, 
we estimate the application range at large $P$ and small $P$, 
corresponding to large temperature 
and small temperature, respectively. 
In the free quark gas limit, where $P$ is maximum, 
the quark determinant is expected to be 
$(\ln \det M)(N_t/N_s)^3 = 
(7 \pi^2/ 60)+(1/2)(\mu_q/T)^2+(1/4\pi^2)(\mu_q/T)^4$
in the continuum limit \cite{BS03}.
Because $\rho_4$ is infinity, the convergence of the Taylor 
expansion seems to be good for large $P$. 

On the other hand, in the study of the equation of state 
\cite{BS05,KRT}, the numerical results of the derivatives of pressure 
with respect to $\mu_q/T$ at low temperature have 
been found to reproduce the prediction from the 
hadron resonance gas model very well.
Because small plaquette values are generated in the low temperature 
simulation, this model may give a suggestion of the application range 
for small $P$.
The quark chemical potential dependence of pressure 
in the hadron resonance gas model is 
discussed in Ref.~\cite{KRT}. It is suggested that
\begin{eqnarray}
\frac{p(\mu_q)}{T^4} - \frac{p(0)}{T^4} 
\propto \cosh \left( \frac{3 \mu_q}{T} \right),
\end{eqnarray}
and the radius of convergence for pressure is given by 
\begin{eqnarray}
\rho_{i}=\sqrt{\frac{(i+2)(i+1)}{9}}.
\end{eqnarray}
This $\rho_i$ increases as $i$ increases, 
and the convergence radius $\rho$ is infinity. 
Although we expect that the radius of convergence for 
$\ln R$ is larger than that for the pressure, 
we try to estimate the application range from this $\rho_i$. 
When we neglect terms higher than $O(\mu_q^6)$, 
as it is done in this study, 
the application range is suggested to be 
$\mu_q/T \simle \rho_6 \approx 2.5$. 
This implies that the error that arises from the approximation 
up to $O(\mu_q^6)$ may be sizeable for $\mu_q/T \sim 2.5$, and 
more careful arguments are required when we calculate 
the reweighting factor $R(P, \mu_q)$ for $\mu_q/T \simge 2.5$.
We will discuss this application range in Sec.~\ref{sec:appli} again.
The results of $\ln R(P, \mu_q)$ obtained by the calculations up to 
$O(\mu_q^4)$ and $O(\mu_q^6)$ will be compared, and we will confirm 
that the truncation error is still small even at $\mu_q/T \sim 2.5$.

\subsection{Avoidance of the sign problem at finite density}
\label{sec:sign}

\begin{figure}[tbp]
\begin{center}
\begin{tabular}{c}
\includegraphics[width=80mm]{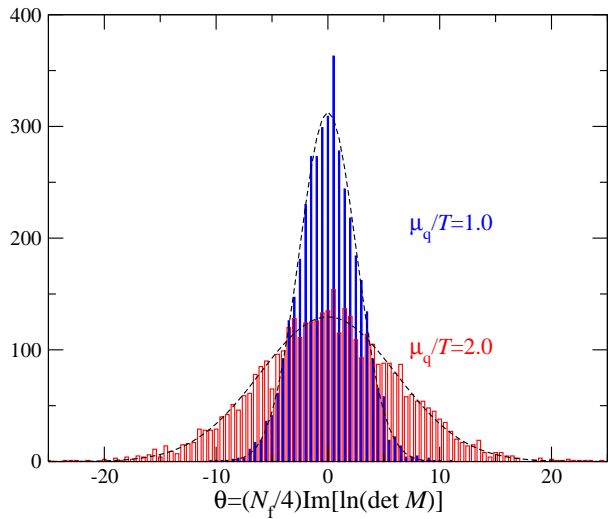}
\end{tabular}
\caption{The histogram of the complex phase for $\mu_q/T=1.0$ 
and $2.0$ at $\beta=3.65$ $(T/T_c=1.00)$ on a $16^3 \times 4$ lattice.
The dashed lines are the fit results by Gaussian functions.}
\label{fig:theta}
\end{center}
\end{figure}

We discuss here how to avoid the sign problem in our 
reweighting approach. 
In the framework of the Taylor expansion, 
we can easily separate $\ln \det M(\mu_q)$ into real and 
imaginary parts because the even derivatives of $\ln \det M(\mu_q)$ 
are real and the odd derivatives are purely imaginary \cite{BS02}. 
The absolute values of the quark determinant and 
the complex phases $\theta$ are thus given by 
\begin{eqnarray}
N_{\rm f} \ln | \det M| &=& N_{\rm f} {\rm Re} \left[ \ln (\det M) \right] 
= N_{\rm f} \sum_{n=0}^{\infty} \frac{1}{(2n)!} 
{\rm Re} \frac{\partial^{2n} (\ln \det M)}{\partial (\mu_q/T)^{2n}} 
\left( \frac{\mu_q}{T} \right)^{2n} , \label{eq:absdet} \\
\theta & = & N_{\rm f} {\rm Im} \left[ \ln (\det M) \right] 
= N_{\rm f} \sum_{n=0}^{\infty} \frac{1}{(2n+1)!} 
{\rm Im} \frac{\partial^{2n+1} (\ln \det M)}{\partial (\mu_q/T)^{2n+1}} 
\left( \frac{\mu_q}{T} \right)^{2n+1} , 
\label{eq:theta}
\end{eqnarray} 
where one must replace $N_{\rm f}$ in these equations to $N_{\rm f}/4$ 
when one uses a staggered type quark action. 
%The comparison between the value of $\theta$ with this approximation and 
%the exact value has been done in Ref.~\cite{dFKT}, and the reliability 
%and the application range have been discussed. 
Here, it is worth noting that $\theta$ corresponds to the complex 
phase of the quark determinant however by definition 
this quantity is not restricted to 
the range from $-\pi$ to $\pi$ because 
there is no reason that the imaginary part of $\ln \det M$ 
in Eq.~(\ref{eq:theta}) must be in the finite range. 
In fact, this quantity becomes larger as the volume increases. 

We show histograms of $\theta$ at the pseudo-critical temperature 
$(\beta=3.65)$ for $\mu_q/T=1.0$ and $2.0$ in Fig.~\ref{fig:theta}, 
where $\theta$ is calculated using the data of the Taylor expansion 
coefficients up to $O(\mu_q^5)$ obtained with two-flavors of 
p4-improved staggered quarks in Ref.~\cite{BS05}.
These histograms seem to be almost Gaussian functions. 
We fit these data by Gaussian functions, $\sim \exp (-x \theta^2)$, 
where the overall factor and $x$ are the fit parameters. 
The dashed lines in Fig.~\ref{fig:theta} are the fit results. 
It is found that the histogram of $\theta$ is well-represented 
by a Gaussian function. 

Similar to the discussion of the Gaussian distribution function 
for the plaquette in Sec.~\ref{sec:tran}, 
we may argue that the histogram of $\theta$ 
is a Gaussian function. 
Because there is no critical point in two-flavor QCD with 
finite quark mass at $\mu_q=0$, the spatial correlation 
length between the quark fields is not expected to be long.
The Taylor expansion coefficients in Eq.~(\ref{eq:theta}) are 
given by combinations of traces of products of 
$\partial^n M/ \partial (\mu_q/T)^n$ and $M^{-1}$ 
(see Appendix of Ref.~\cite{BS05}). 
Therefore, the expansion coefficients are obtained by the sum 
of the diagonal elements of such matrices. 
When the correlation among the diagonal elements is small and 
the volume is sufficiently large, the distribution functions of 
the expansion coefficients and $\theta$ should be of Gaussian type 
due to the central limit theorem. 
For example, the diagonal elements of the first coefficient, 
${\rm Im} [\partial (\ln \det M)/ \partial (\mu_q/T)]= 
{\rm Im} [{\rm tr} [M^{-1} (\partial M/ \partial (\mu_q/T))]]$, is 
the imaginary part of the local number density operator at $\mu_q=0$. 
If the spatial density correlation is not very strong, 
the Gaussian distribution is expected. 
Figure \ref{fig:theta} is consistent with this argument. 

%Moreover, we can also define the complex phase of the quark determinant 
%$\theta = N_{\rm f} {\rm Im} \left[ \ln (\det M) \right] $ 
%by the sum of complex phases of the eigenvalues, i.e. 
%$\ln \det M = \sum_n \ln (|\lambda_n| e^{i \theta_n})
%=\sum_n \ln |\lambda_n| +i \sum_n \theta_n$, denoting 
%the eigenvalues $|\lambda_n| e^{i \theta_n}$. 
%This definition is different from the definition of $\theta$ we use 
%in this study, but we may argue the Gaussian distribution applying 
%the central limit theorem. 
%Although the eigenvalues correlate with each other, if the eigenvalues 
%can be classified into some independent groups 
%and the number of the groups becomes larger as the volume increases, 
%the distribution function of $\theta$ should be of Gaussian 
%in the infinite volume limit.

We note that, once we assume a Gaussian distribution for $\theta$, 
the problem of complex weights can be avoided.
A variety of distribution functions with respect to various quantities 
are discussed in the density of state method 
\cite{Swen88,Gock88,JN01,Ta04,FKS07}. 
We introduce the probability distribution $\bar{w}$ as 
a function of the plaquette $P$, the absolute value of 
$[\det M(\mu_q)/ \det M(0)]^{N_{\rm f}} \equiv F$ 
and the complex phase $\theta \equiv {\rm Im} [ \ln F(\mu_q)]$, 
\begin{eqnarray}
\bar{w}(P', |F|', \theta') \equiv 
\int {\cal D}U \delta (P'-P) \delta (|F|'-|F|)
\delta (\theta' - \theta) 
(\det M(0))^{N_{\rm f}} e^{6\beta N_{\rm site} P}.
\end{eqnarray}
The distribution function itself is defined as an expectation value 
at $\mu_q=0$, i.e. 
$\bar{w}(P', |F|', \theta') \propto 
\langle \delta (P'-P) \delta (|F|'-|F|)
\delta (\theta'- \theta) \rangle_{(T, \mu_q=0)}$, 
however $|F|$ and $\theta$ are functions of $\mu_q/T$ obtained by 
the Taylor expansion at $\mu_q=0$. 
The expectation value of ${\cal O}[P, |F|, \theta]$ 
at $\mu_q=0$ is given by 
\begin{eqnarray}
\left\langle {\cal O}[P, |F|, \theta] \right\rangle_{(T, \mu_q=0)}
= \frac{1}{{\cal Z}(\mu_q=0)}\int d P \int d|F| \int d \theta \ 
\ {\cal O}[P, |F|, \theta] \ \bar{w}(P, |F|, \theta). 
\label{eq:apdist}
\end{eqnarray}

Since the partition function is real even at non-zero density, 
the distribution function has the symmetry under 
the change from $\theta$ to $-\theta$.
Therefore, the distribution function is a function of $\theta^2$, 
e.g., $\bar{w}(\theta) \sim \exp[-(a_2 \theta^2 +a_4 \theta^4 
+a_6 \theta^6 + \cdots)].$
Moreover, as we discussed, when the system size is sufficiently 
large in comparison to the correlation length, 
the distribution function should be well-approximated 
by a Gaussian function:
\begin{eqnarray}
\bar{w}(P, |F|, \theta) \approx \sqrt{\frac{a_2 (P, |F|)}{\pi}} 
\bar{w}'(P, |F|) \exp \left[-a_2 (P, |F|) \theta^{2} \right].
\label{eq:phap}
\end{eqnarray}
We assume this distribution function in terms of
$\theta$ when $P$ and $|F|$ are fixed.

The coefficient $a_2 (P, |F|)$ is given by 
\begin{eqnarray}
\frac{1}{2a_2 (P', |F|')} &=&
\left. \int d \theta \ \theta^2 \ \bar{w}(P', |F|', \theta) \right/
\int d \theta \ \bar{w}(P', |F|', \theta)
\nonumber \\
%&=&
%\frac{ \int {\cal D}U  \ \theta^2 \ \delta (P'-P) 
%\delta (|F|'-|F|) (\det M(0))^{N_{\rm f}} e^{-S_g} }{
%\int {\cal D}U \delta (P'-P) \delta (|F|'-|F|)
%(\det M(0))^{N_{\rm f}} e^{-S_g}}
%\nonumber \\
&=&\frac{ \left\langle \theta^2 \delta (P'-P) 
\delta (|F|'-|F|) \right\rangle_{(T, \mu_q=0)} }{ \left\langle 
\delta (P'-P) \delta (|F|'-|F|) 
\right\rangle_{(T, \mu_q=0)}},
\label{eq:a2}
\end{eqnarray}
using 
$ \sqrt{a_2/\pi} \int \theta^2 \exp(-a_2 \theta^2) d \theta 
= 1/(2a_2)$. 
%This equation is obtained from Eqs.~(\ref{eq:apdist}), (\ref{eq:phap}) and 
%\begin{eqnarray}
%\sqrt{\frac{a_2}{\pi}} \int \theta^2 e^{-a_2 \theta^2} d \theta 
%= -\sqrt{\frac{a_2}{\pi}} \frac{d \sqrt{\pi/a_2}}{da_2} = \frac{1}{2a_2}. 
%\end{eqnarray}

When the volume is sufficiently large, this assumption will be valid 
except at a critical point.
For the case of two-flavor QCD at finite quark mass, this assumption 
should be valid because there is no critical point for $\mu_q=0$ except 
in the chiral limit, and is suggested by Fig.~\ref{fig:theta}, 
though the values of $P$ and $|F|$ are not fixed in the calculation 
of Fig.~\ref{fig:theta}.
Then, the integration over $\theta$ can be carried out easily and we obtain 
the numerator of Eq.~(\ref{eq:rmu}) for the calculation of 
$R(P, \mu_q)$, 
\begin{eqnarray}
\left\langle F(\mu_q) \delta(P'-P) \right\rangle_{(T, \mu_q=0)}
&\approx& \frac{1}{\cal Z} \int dP \int d|F| \int d \theta 
\sqrt{\frac{a_2}{\pi}} \bar{w}'(P, |F|) e^{-a_2 \theta^2} 
e^{i \theta} |F| \delta(P'-P) \nonumber \\
%&=& \int dP \int d|F| \int d \theta 
%\sqrt{\frac{a_2}{\pi}} \ \bar{w}'(P, |F|) e^{-a_2 (\theta -i/(2a_2))^2}
%e^{-1/(4a_2)} |F| \delta(P-P') \nonumber \\
&=& \frac{1}{\cal Z} \int dP \int d|F| \ \bar{w}'(P, |F|) 
e^{-1/(4a_2)} |F| \delta(P'-P) \nonumber \\
&=& \frac{1}{\cal Z} \int {\cal D}U e^{-1/(4a_2(P, |F|))} |F(\mu_q)| 
\delta(P'-P) (\det M(0))^{N_{\rm f}} e^{-S_g} \nonumber \\
&=& \left\langle e^{-1/(4a_2(P, |F|))} 
| F(\mu_q) | \delta(P'-P) \right\rangle_{(T, \mu_q=0)}.
\label{eq:den12}
\end{eqnarray}
Since $\theta$ is roughly proportional to the size of the quark 
matrix $M$, the value of $1/a_2$ becomes larger as the 
volume increases. Therefore, the phase factor in $R(P,\mu_a)$ decreases 
exponentially as a function of the volume. 
However, the most important point in this approach is that the operator 
in Eq.~(\ref{eq:den12}) is always real and positive for each configuration 
in this framework, hence the expectation value of $R(P, \mu_q)$ 
is always larger than its statistical error, i.e. 
the contribution $\ln R(P,\mu_q)$ to the effective potential 
$V(P, \beta, \mu_q)$ is always well-defined.
Therefore, the sign problem is completely avoided if we can assume 
the Gaussian distribution of $\theta$. 

We calculate $\theta$ using the stochastic noise method. 
Then, the value of $\theta$ contains an error due to the finite 
number of noise vectors $(N_{\rm n})$. 
As discussed in Ref.~\cite{BS02,ej04}, a careful treatment is required to 
reduce this error for the calculation of $\sqrt{\langle \theta^2 \rangle}$, 
i.e. width of the distribution of $\theta$.
Since the noise sets for the calculation of the two $\theta$ in the 
product must be independent, we subtract the contributions from using 
the same noise vector for each factor. 
By using this method, we can make the $N_{\rm n}$-dependence of 
$\sqrt{\langle \theta^2 \rangle}$ much smaller than that by the naive 
calculation from rather small $N_{\rm n}$, 
hence it may be closer to the $N_{\rm n}=\infty$ limit. 
We took $N_{\rm n}=50$ or $100$ in this calculation, so that 
the $N_{\rm n}$-dependence is negligible.
On the other hand, as $N_n$ increases, the result of 
$\sqrt{\langle \theta^2 \rangle}$ obtained by the naive calculation 
without the subtraction becomes smaller and 
approaches the result with the subtraction. 
For the case at $\beta=3.65$, $\mu_q/T=2.0$ with $N_n=100$ in 
Fig.~\ref{fig:theta}, the difference between them is about $13 \%$. 
Since the width of the distribution function shown in 
Fig.~\ref{fig:theta} corresponds to $\sqrt{\langle \theta^2 \rangle}$ 
without the subtraction, the width in Fig.~\ref{fig:theta} is slightly 
larger than that in the $N_{\rm n}=\infty$ limit. 

\begin{figure}[tbp]
\begin{center}
\begin{tabular}{c}
\includegraphics[width=80mm]{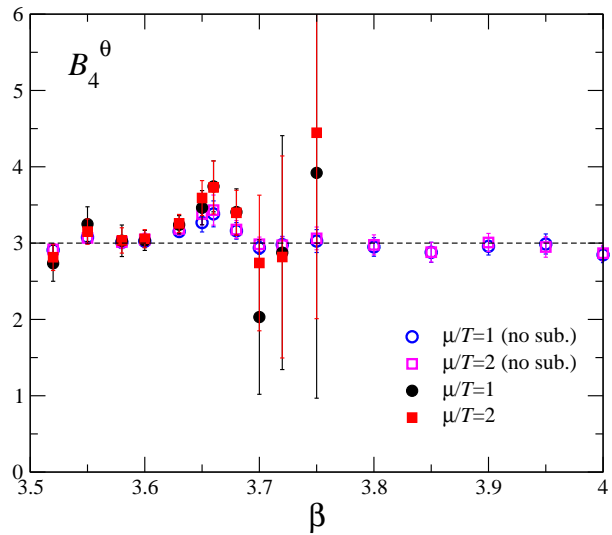}
\end{tabular}
\caption{The fourth order Binder cumulant of the complex phase for 
$\mu_q/T=1.0$ (circle) and $2.0$ (square).
The filled symbols are the results obtained when the contributions 
from using the same noise vector are subtracted in the products of $\theta$. 
The open symbols are the results without the subtraction.
The dashed line is the value of Gaussian distribution.}
\label{fig:b4theta}
\end{center}
\end{figure}

For more quantitative arguments of the Gaussian distribution function, 
we also compute the fourth order Binder cumulant of the complex phase 
for $\mu_q/T=1.0$ and $2.0$, using the data obtained in a simulation 
of two-flavor QCD with p4-improved staggered quarks, Ref.~\cite{BS05}. 
The Binder cumulant is defined by 
\begin{eqnarray}
B_4^{\theta} \equiv 
\frac{\left\langle \theta^4 \right\rangle}
{\left\langle \theta^2 \right\rangle^2}. 
\label{eq:b4theta}
\end{eqnarray}
As discussed in Sec.~\ref{sec:tran}, this quantity is a good indicator 
to check whether the distribution is of Gaussian or not.
To confirm the validity of the assumption, Eq.~(\ref{eq:phap}), 
we should compute $B_4^{\theta}$ as a function of $P$ and $|F|$. 
However, because the width of the plaquette distribution function 
$w(P,\beta)$ for each $\beta$ is narrow in our simulation 
(see the results in Sec.~\ref{sec:rewbeta}), 
we calculate $B_4^{\theta}$ for each $\beta$ without 
separating the configurations in terms of $P$.
The circle and square symbols in Fig.~\ref{fig:b4theta} 
are the results for $\mu_q/T=1.0$ and $2.0$, respectively. 
We use the stochastic noise method for the calculation of the products 
of $\theta$. The results plotted by filled symbols are obtained when 
the contributions from using the same noise vector are subtracted.
The open symbols are the results without the subtraction.
Because the complex phase vanishes in the large $\beta$ limit, 
$\langle \theta^2 \rangle$ becomes smaller as $\beta$ increases. 
We omitted results having large statistical errors due to the small 
$\langle \theta^2 \rangle$ at large $\beta$. 
We find from this figure that the results of $B_4^{\theta}$ are almost 
consistent with three. 
As we discussed in Sec.~\ref{sec:tran}, if the distribution is of 
Gaussian, the Binder cumulant is three. 
Hence, this figure suggests the Gaussian distribution.
The results around $\beta = 3.66$ are slightly larger than three, 
but the difference would be within the systematic error due to 
finite statistics because usually the Binder cumulant becomes 
smaller than three when the correlation length is long.

To estimate the effect when the distribution is slightly different from 
Gaussian, we consider a distribution function with small $a_4$, i.e. 
$\bar{w}(\theta) \sim \exp[-a_2 \theta^2 -a_4 \theta^4]$. 
In this case, the phase factor changes from $\exp[-1/(4a_2)]$ to 
$\exp[-1/(4a_2) +3a_4/(4a_2^3) -a_4/(16a_2^4)+ \cdots]$, 
and also the expectation value of $\theta^2$ for fixed $P$ and $|F|$ 
becomes $\langle \theta^2 \rangle = 1/(2a_2) -3a_4/(2a_2^3) +\cdots$. 
Since the term of $3a_4/(4a_2^3)$ is absorbed into 
$\langle \theta^2 \rangle /2$, the leading contribution from $a_4$ 
in the phase factor is $\exp[-a_4/(16a_2^4)]$. 
Because $1/a_2 \sim O(\mu_q^2)$, this effect becomes 
larger as $\mu_q$ increases. 
Therefore, for the case of $a_4 \neq 0$ $(B_4^{\theta} \neq 3)$, 
the estimation of the range of $\mu_q$ in which the non-Gaussian 
contribution is small may be important as well as the application range 
of the Taylor expansion in $\mu_q$ discussed in the previous section.

\subsection{Reweighting method for $\beta$ direction}
\label{sec:rewbeta}

\begin{figure}[tbp]
\begin{center}
\begin{tabular}{c}
\includegraphics[width=80mm]{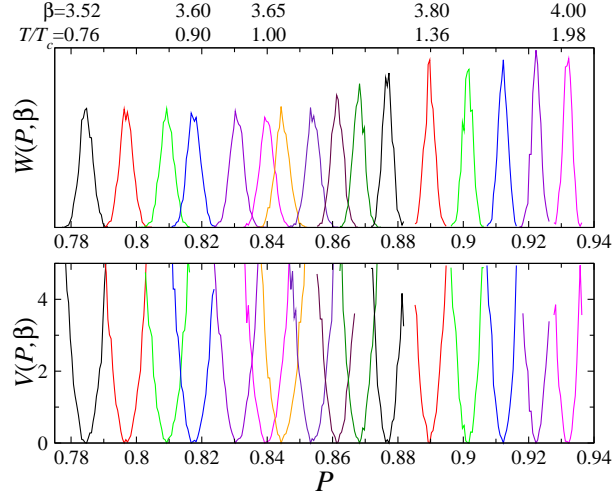}
\end{tabular}
\caption{The plaquette histogram and the effective potential at 
$\mu_q=0$ as a function of the plaquette for the two-flavor 
p4-improved staggered action obtained in Ref.~\cite{BS05}.}
\label{fig:hist}
\end{center}
\end{figure}

\begin{figure}[tbp]
\begin{center}
\begin{tabular}{c}
\includegraphics[width=80mm]{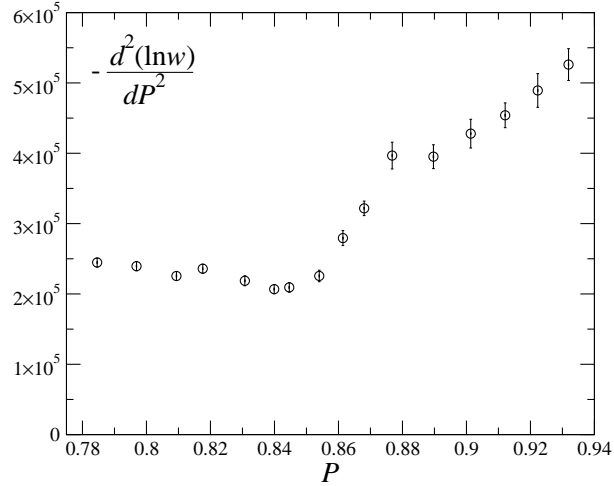}
\end{tabular}
\caption{The curvature of the effective potential at $\mu_q=0$.}
\label{fig:d2vdp2}
\end{center}
\end{figure}

We consider the reweighting method for the $\beta$ 
direction at $\mu_q=0$. This is the case of $R(P,0)=1$. 
Using the plaquette distribution function (plaquette histogram) 
$w(P, \beta_0)$ at the simulation point $\beta_0$,  
the expectation value of an operator given by the plaquette 
is evaluated by 
\begin{eqnarray}
\langle {\cal O}[P] \rangle (\beta) 
= \frac{\int {\cal O}[P] 
e^{6 (\beta - \beta_0) N_{\rm site} P} w(P,\beta_0) \ dP}
{\int e^{6 (\beta - \beta_0) N_{\rm site} P} w(P,\beta_0) \ dP}, 
\label{eq:rewbeta}
\end{eqnarray}
where we discuss only the case when the quark matrix does not depend 
on $\beta$ explicitly for simplicity, otherwise equation 
(\ref{eq:rewbeta}) is no longer correct.

From Eq.~(\ref{eq:rewbeta}), under the parameter change from 
$\beta_0$ to $\beta$, the weight $w(P, \beta)$ becomes 
\begin{eqnarray}
w(P, \beta) = e^{6 (\beta - \beta_0) N_{\rm site} P} w(P, \beta_0). 
\label{eq:weibeta}
\end{eqnarray} 
If we rewrite 
$e^{-6 \beta_0 N_{\rm site} P} w(P, \beta_0) \equiv w(P)$,
we obtain Eq.~(\ref{eq:rewmuP}) from Eq.~(\ref{eq:rewmu}). 
The effective potential becomes
\begin{eqnarray}
V(P,\beta)= -\ln w(P,\beta) 
= V(P,\beta_0) -6 (\beta - \beta_0) N_{\rm site} P.
\label{eq:vrewbeta}
\end{eqnarray}
When $\beta$ is increased, the slope of $V(P)$ becomes smaller, 
whereas the curvature of $V(P)$ does not change. 
This implies that the curvature of $V(P)$ is independent of $\beta$. 
For the case of $d^2V/dP^2>0$, the value of $P$ 
which gives the minimum of $V(P)$ becomes larger as $\beta$ increases. 

Here, we want to explain the $\beta$ dependence of the 
effective potential using the data from Ref.~\cite{BS05}. 
The configurations were generated with Symanzik-improved 
gauge and two-flavor p4-improved staggered fermion actions.
% on a $16^3 \times 4$ lattice. 
Because the improved gauge action was used in Ref.~\cite{BS05}, 
the definition of $P$ is 
\begin{eqnarray}
P=-\frac{S_g}{6 N_{\rm site} \beta} 
= \frac{1}{6 N_{\rm site}}  \left\{ 
\frac{5}{3}\sum_{x,\, \mu > \nu} W_{\mu \nu}^{1 \times 1}(x) 
-\frac{1}{12}\sum_{x,\, \mu \neq \nu} W_{\mu \nu}^{1 \times 2}(x) \right\},
\end{eqnarray}
where $W_{\mu \nu}^{I \times J}$ is the $I \times J$ Wilson loop 
for each point and each direction.
The maximum of this $P$ is 1.5.

The probability distribution function $w(P)$, i.e. the histogram 
of $P$, and the effective potential $V(P)$ are given in Fig.~\ref{fig:hist}. 
These are measured at sixteen simulation points from $\beta=3.52$ 
to $4.00$ for the bare quark mass $ma=0.1$. 
The corresponding temperature normalized by the pseudo-critical 
temperature is in the range of $T/T_c= 0.76$ to $1.98$, 
and the pseudo-critical point $(T/T_c=1)$ is about $\beta=3.65$. 
We show the values of $\beta$ and $T/T_c$ above these figures. 
The ratio of pseudo-scalar and vector meson masses is 
$m_{\rm PS}/m_{\rm V} \approx 0.7$ at $\beta=3.65$.
The lattice size $N_{\rm site}$ is $16^3 \times 4$. 
The number of configurations is 1000 -- 4000 for each $\beta$. 
Further details on the simulation parameters are given 
in Ref.~\cite{BS05}.
To obtain $w(P)$ and $V(P)$, 
we grouped the configurations by the value of $P$ into blocks 
and counted the number of configurations in these blocks, 
and the potential $V(P)$ is normalized by the minimum value 
for each temperature.

Because the transition from the hadron phase to the quark-gluon 
plasma phase is a crossover transition for two-flavor QCD 
with finite quark mass, 
the distribution function is always of Gaussian type, 
i.e. the effective potential is always a quadratic function. 
The value of the plaquette at the potential minimum increases
as $\beta$ increases in accordance with the above argument.

Figure \ref{fig:d2vdp2} shows the curvature of the effective 
potential at $\mu_q=0$, $d^2V/dP^2 (P) = -d^2(\ln w)/dP^2$,  
as a function of $P$. 
We estimate this quantity from the relation between 
the plaquette susceptibility $\chi_P$ and 
the curvature of the potential at $\mu_q=0$, Eq.~(\ref{eq:curpot}).
Here, it should be emphasized again that the slope of the potential 
changes as Eq.~(\ref{eq:vrewbeta}) when $\beta$ is changed, 
but the curvature of the potential never changes. 
This means that the curvature is 
independent of $\beta$ and is determined by the measure 
${\cal D} U$ and the quark determinant $\det M$.
As we discussed in Sec.~\ref{sec:tran}, the curvature of 
the effective potential $V(P)$ at $P$ for the potential minimum 
is important to categorize the nature of phase transition, e.g. 
the curvature must be zero at a second order phase transition point. 
The property of the curvature being independent of $\beta$ will 
make our analysis simpler in the next section.

\subsection{Numerical calculations of the reweighting factor}
\label{sec:numerical}

\begin{figure}[tbp]
\begin{center}
\begin{tabular}{c}
\includegraphics[width=80mm]{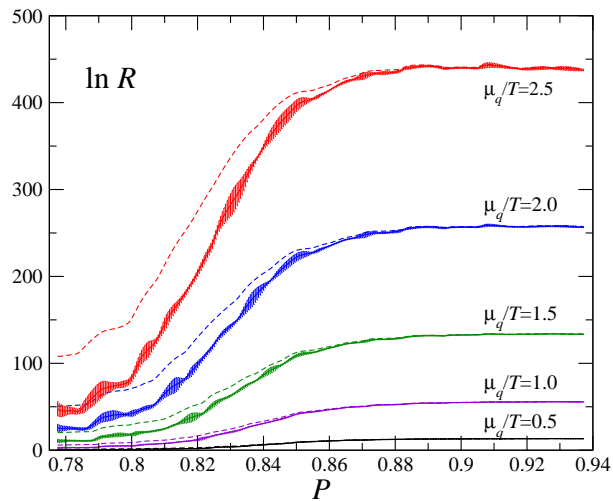}
\end{tabular}
\caption{The reweighting factor $R(P, \mu_q)$ for $\mu_q/T=0.5$ -- $2.5$ 
obtained by the Taylor expansion up to $O(\mu_q^6)$. 
The dashed lines are the cases when the effect of the complex phase 
is omitted, $\bar{R}(P, \mu_q)$.}
\label{fig:rmu}
\end{center}
\end{figure}

We calculate the probability distribution function at non-zero $\mu_q$ 
using the data of the Taylor expansion coefficients up to $O(\mu_q^6)$ 
computed in Ref.~\cite{BS05} with the p4-improved staggered quark action. 
Since the simulations are performed in the region where no critical 
points exist, the assumption of the Gaussian function is valid. 
The coefficient $a_2 (P, |F|)$ in the distribution function of 
$\theta$ is measured using Eq.~(\ref{eq:a2}). 
However, because the values $P$ and 
$|F|=|\det M(\mu_q) /\det M(0)|^{N_{\rm f}}$ on each configuration 
are strongly correlated \cite{ej04}, 
we may assume that $|F|$ is approximately given as a function of 
$P$ for each configuration so that $a_2 (P, |F|)$ 
is given by a function of $P$ only. 
In this approximation, the contribution from the complex phase 
in $R(P', \mu_q)$ can be simplified, 
\begin{eqnarray}
R(P', \mu_q) \approx e^{-1/(4a_2(P'))} \frac{
\left\langle | F(\mu_q) | \delta(P'-P) \right\rangle_{(T, \mu_q=0)}}{
\left\langle \delta(P'-P) \right\rangle_{(T, \mu_q=0)}}. 
\end{eqnarray}
Although the correlation between $|F|$ and $a_2$ is neglected in 
this equation, the main contribution to the variation of $R(P, \mu_q)$ 
comes from $|F|$, and the contribution from the phase factor 
is not very large, as we will see in Fig.~\ref{fig:rmu}. 
Therefore, the correlation of these two factors is negligible 
in the following argument.
For the calculation of $R(P, \mu_q)$, 
we use the delta function approximated by a Gaussian function,
$\delta(x) \approx 1/(\Delta \sqrt{\pi}) \exp[-(x/\Delta)^2]$, 
where $\Delta=0.0025$ is adopted. 

Because $R(P, \mu_q)$ is independent of $\beta$, we mix all data 
obtained at different $\beta$. This mixture can be justified by 
extending Eq.~(\ref{eq:rmu}) for multi-$\beta$, e.g. 
$R(P', \mu_q)=[N_1 \langle \delta(P'-P) F \rangle_{\beta_1} 
+ N_2 \langle \delta(P'-P) F \rangle_{\beta_2}]/
[N_1 \langle \delta(P'-P) \rangle_{\beta_1} 
+ N_2 \langle \delta(P'-P) \rangle_{\beta_2}]$
for the data at $\beta_1$ and $\beta_2$ with the number of configurations 
$N_1$ and $N_2$.
The results for $\ln R(P, \mu_q)$ are shown by solid lines in Fig.~\ref{fig:rmu} 
for $\mu_q/T=0.5, 1.0, 1.5, 2.0$ and $2.5$. 
We find a rapid change in $\ln R$ around $P \sim 0.83$, and 
the variation becomes larger as $\mu_a/T$ increases. 

The dashed lines in Fig.~\ref{fig:rmu} are the results that we obtained 
when the effect of the complex phase,
i.e. $\exp[-1/(4a_2)]$, is omitted. We define this quantity as 
\begin{eqnarray}
\bar{R}(P', \mu_q) \equiv \frac{
\left\langle | F(\mu_q) | \delta(P'-P) \right\rangle_{(T, \mu_q=0)}}{
\left\langle \delta(P'-P) \right\rangle_{(T, \mu_q=0)}} .
\end{eqnarray}
We discuss in Sec.~\ref{sec:isospin} that 
these dashed lines correspond to the reweighting 
factor with non-zero isospin chemical potential $\mu_I$ and 
zero quark chemical potential $\mu_q$, i.e. 
$\bar{R}(P, \mu_q) = R(P, \mu_I).$ 
The variation of $\ln R$ in terms of $P$ becomes milder 
when the effect of the complex phase is omitted. 

The effective potential $V(P, \beta, \mu_q)$ is obtained from 
Eq.~(\ref{eq:potmu}) substituting the data in 
Fig.~\ref{fig:hist} and Fig.~\ref{fig:rmu}. 
To study the existence of a second order phase transition, 
the curvature of the potential is important. 
The minimum of the potential can be changed by shifting $\beta$ but 
the curvature can be controlled only by $\ln R(P, \mu_q)$. 
The result of the curvature at $\mu_q=0$, $-d^2(\ln w)/dP^2$, 
as a function of $P$ is shown in Fig.~\ref{fig:d2vdp2}. 
%which is estimated by Eq.~(\ref{eq:gcoe}). 
Because $-d^2(\ln w)/dP^2$ is positive, a region where 
$d^2(\ln R)/dP^2 > 0$ is necessary for the existence of a critical 
point. The curvature of $\ln R$ is positive for $P \simle 0.83$.

\begin{figure}[tbp]
\begin{center}
\begin{tabular}{c}
\includegraphics[width=80mm]{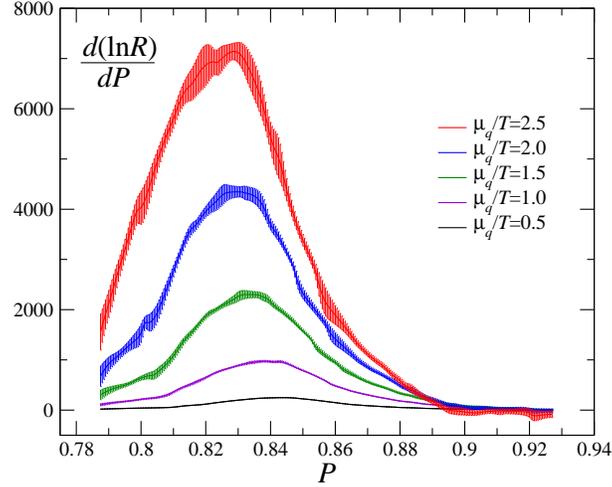}
\end{tabular}
\caption{The slope of $\ln R (P, \mu_q)$ as functions of the plaquette.}
\label{fig:potslp}
\end{center}
\end{figure}

\begin{figure}[tbp]
\begin{center}
\begin{tabular}{c}
\includegraphics[width=80mm]{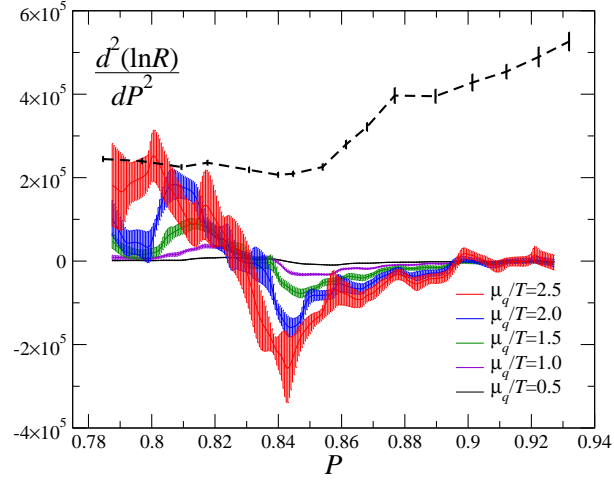}
\end{tabular}
\caption{The curvature of $\ln R (P, \mu_q)$ as functions of the plaquette.
The dashed line is the curvature of $-\ln w$. }
\label{fig:potcur}
\end{center}
\end{figure}

In order to analyze the sign of $d^2V/dP^2(P, \mu_q)$, 
we fitted the data around $P$ by a quadratic function, 
$\ln R(P', \mu_q) = x_0 +x_1 (P'-P) +x_2 (P'-P)^2$, where 
$x_0,x_1$ and $x_2$ are the fit parameters, and 
calculate the first and second derivatives of $\ln R(P, \mu_q)$ 
at each $P$. 
The result of the slope, $d (\ln R)/dP (P, \mu_q)=x_1$, is 
shown in Fig.~\ref{fig:potslp} for each $\mu_q/T$. 
We adopt the result obtained by fitting in the range 
between $P -0.015$ and $P +0.015$ for each $P$ as the final result.
In the region around $P \sim 0.83$, $d (\ln R)/dP$ 
becomes larger as $\mu_q/T$ increases and $\ln R(P, \mu_q)$ 
changes sharply in this region.
The result of the curvature, $d^2 (\ln R)/dP^2 (P, \mu_q)=2x_2$, is 
plotted as solid line in Fig.~\ref{fig:potcur}. 
The magnitude of the curvature of $\ln R$ also becomes larger 
as $\mu_q/T$ increases.
The dashed line in Fig.~\ref{fig:potcur} is the data of 
$-d^2 (\ln w)/dP^2(P)$ in Fig.~\ref{fig:d2vdp2}.
This figure indicates that the maximum value of 
$d^2 (\ln R)/dP^2 (P, \mu_q)$ at $P=0.80$ becomes larger 
than $-d^2 (\ln w)/dP^2$ for $\mu_q/T \simge 2.5$. 
This suggests that the curvature of the effective potential, 
$d^2 V/dP^2 =-d^2(\ln w)/dP^2 -d^2 (\ln R)/dP^2$, vanishes at 
$\mu_q/T \sim 2.5$ and a region of $P$ where the curvature is 
negative appears for large $\mu_q/T$.

Next, we estimate the value of $\beta$ which gives 
the potential minimum at $P=0.8$ for $\mu_q/T=2.5$
by solving the equation: 
\begin{eqnarray}
\frac{dV}{dP} (P, \beta, \mu_q) 
= -\frac{d(\ln R)}{dP} (P, \mu_q) - \frac{d(\ln w)}{dP} (P, \beta_0)
 -6(\beta - \beta_0)N_{\rm site} =0. 
\end{eqnarray}
This equation can be solved without changing $\mu_q/T$, and tells us 
the location of the critical point in the $(\beta, \mu_q/T)$ plane.
Since a simulation with $\beta \approx 3.56$ 
gives $d(\ln w)/dP=0$ at $P \approx 0.8$, we adopt $\beta_0=3.56$. 
Substituting $d (\ln R)/dP \approx 4000$ at $(P, \mu_q/T) =(0.8, 2.5)$ 
in Fig.~\ref{fig:potslp} and $N_{\rm site}=16^3 \times 4$, 
we obtain $\beta \approx 3.52$.
This $\beta$ corresponds to $T/T_c=0.76$, where $T_c$ is the 
pseudo-critical temperature at $\mu_q=0$.
Therefore, it is found that the potential is flat 
up to second order in $P$ around $P=0.80$ with 
$(T/T_c, \mu_q/T) \approx (0.76, 2.5)$, suggesting 
the existence of a critical point around this value. 

Further studies are, of course, needed for the precise 
determination of the critical point in the $(T, \mu_q)$ plane, 
increasing the number of terms in the Taylor expansion of 
$\ln \det M$ and decreasing the quark mass in the simulation. 
The quark mass is still heavier than the physical quark mass.
However, the arguments given above indicate the existence of 
a first order phase transition line at large $\mu_q/T$ 
because the magnitude of the curvature of $R(P, \mu_q)$ 
increases monotonically and eventually the curvature of 
the potential becomes negative at large $\mu_q/T$, 
corresponding to a double-well potential of 
a first order phase transition.

\subsection{Application range of this analysis}
\label{sec:appli}

\begin{figure}[tbp]
\begin{center}
\begin{tabular}{c}
\includegraphics[width=80mm]{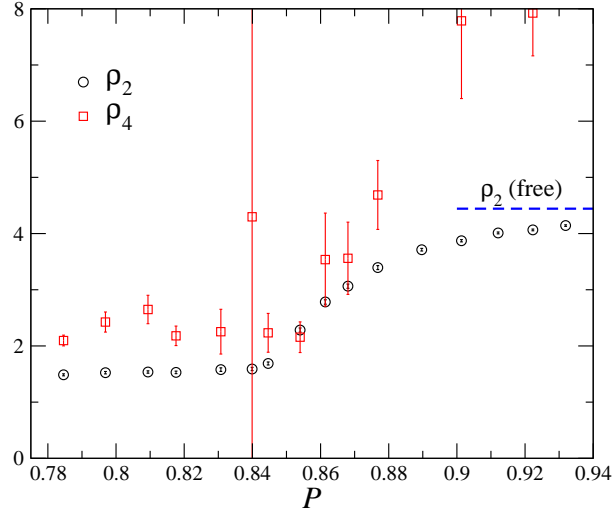}
\end{tabular}
\caption{The radius of convergence, $\rho_2, \rho_4$, 
for the Taylor expansion of $\bar{R}(P, \mu_q)$.}
\label{fig:radcon}
\end{center}
\end{figure}

\begin{figure}[tbp]
\begin{center}
\begin{tabular}{c}
\includegraphics[width=80mm]{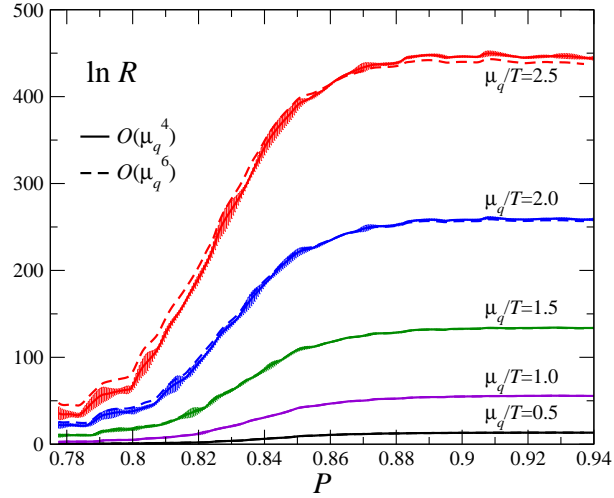}
\end{tabular}
\caption{The reweighting factor $R(P, \mu_q)$ computed by the 
Taylor expansion up to $O(\mu_q^4)$ (solid lines) and 
$O(\mu_q^6)$ (dashed lines).}
\label{fig:trerr}
\end{center}
\end{figure}

Next, we discuss the reliability of our analysis 
in view of the truncation of the Taylor expansion used here. 
Because the dominant contribution in $\ln R$ is given by 
the reweighting factor without the phase effect, $\ln \bar{R}$, 
we consider the radius of convergence for $\ln \bar{R}$. 
The expansion is defined by 
\begin{eqnarray}
\ln \bar{R}(P, \mu_q) &=& 
\sum_{n=1}^{\infty} \bar{r}_n (P) \left( \frac{\mu_q}{T} \right)^n, \\
&& \hspace{-2cm} 
\bar{r}_2 = \langle d_2 \rangle_P, \hspace{5mm}
\bar{r}_4 = \langle d_4 \rangle_P + \frac{1}{2} \left( 
\langle d_2^2 \rangle_P - \langle d_2 \rangle_P^2 \right), 
\nonumber \\ && \hspace{-2cm} 
\bar{r}_6 = \langle d_6 \rangle_P + \langle d_2 d_4 \rangle_P 
- \langle d_2 \rangle_P \langle d_4 \rangle_P
+ \frac{1}{6} \left( \langle d_2^3 \rangle_P -3 \langle d_2 \rangle_P 
\langle d_2^2 \rangle_P +2 \langle d_2 \rangle_P^3 \right), 
\end{eqnarray}
where 
$d_n= (N_{\rm f}/n!) \partial^n (\ln \det M)/\partial (\mu_q/T)^n$, 
$\langle \cdots \rangle_{P'} = \langle \cdots \delta(P'-P) \rangle 
/ \langle \delta(P'-P) \rangle$, 
and the odd terms are zero. 
The radius of convergence is obtained by analyzing 
the asymptotic behavior of 
$\rho_{n}=\sqrt{|\bar{r}_{n}/\bar{r}_{n+2}|}$ for 
$n=2,4,6,\cdots, \infty$.

In this analysis, we calculated $\ln \det M$ using the data of 
$d_n$ up to $O(\mu_q^6)$. This approximation does not affect 
the calculations of $\bar{r}_2, \bar{r}_4$ and $\bar{r}_6$, 
but there is a missing term, i.e. $\langle d_8 \rangle_P$, 
in the calculation of $\bar{r}_8$. 
If the $8^{\rm th}$ order term of $\ln R$ is larger than the 
$6^{\rm th}$ order term, the effect of the truncation may be sizeable.
Because $|\bar{r}_6 (\mu_q/T)^6| > |\bar{r}_8 (\mu_q/T)^8|$
for $\mu_q/T < \rho_6$, 
the application range for our current analysis 
should be $\mu_q/T \simle \rho_6$. 
We calculate $\rho_2$ and $\rho_4$. 
These results are shown in Fig.~\ref{fig:radcon}. 
The dashed line is $\rho_2$ in the free gas limit, and  
$\rho_4$ is infinity in the free gas limit. 
The results of $\bar{r}_2$ and $\bar{r}_4$ are positive 
for all $P$ we investigated, but $\bar{r}_6$ changes its sign 
at $P=0.84$. $\bar{r}_6$ is negative for $P \ge 0.84$. 
We find that $\rho_4$ (square) is larger than $\rho_2$ (circle), 
and the values of $\rho_2$ and $\rho_4$ are larger than 
the hadron resonance gas model values, $\rho_2 \approx 1.15$ 
and $\rho_4 \approx 1.83$. 
For our analysis, where we omitted the calculation of $d_n$ 
higher than the $6^{\rm th}$ order in $\ln R$, the application range 
given by $\rho_6$ would be larger than the hadron resonance gas 
model prediction, $\rho_6 \approx 2.49$, and the parameter range 
we investigated thus seems to be within the application range.

We moreover estimate the effect from higher order terms in the Taylor 
expansion by changing the number of terms in the Taylor expansion.
Figure \ref{fig:trerr} shows the difference between 
the results up to $O(\mu_q^4)$ and $O(\mu_q^6)$ 
for $\mu_q/T=0.5, 1.0, 1.5, 2.0$ and $2.5$. 
The dashed lines are the same as the solid lines in 
Fig.~\ref{fig:rmu} and the solid lines are the results 
obtained when the highest order term and the next 
highest order term, $d^6 (\ln M)/d(\mu_q/T)^6$ and 
$d^5 (\ln M)/d(\mu_q/T)^5$, are omitted in Eq.~(\ref{eq:detTay}).
It is found from this figure that the difference becomes 
visible at $\mu_q/T \sim 2.5$, but the truncation error 
of the Taylor expansion does not affect the qualitative 
argument of the effective potential at finite density 
in the range we have discussed. 
For more quantitative investigation of the critical 
point in the $(T, \mu_q)$ plane, more accurate calculations 
including higher order terms in the Taylor expansion of 
$\mu_q$ may be important.

\subsection{QCD with an isospin chemical potential}
\label{sec:isospin}

Finally, it is worth discussing the difference between QCD with 
a quark (baryon) chemical potential and an isospin chemical potential. 
The isospin chemical potential is defined by 
$\mu_I = (\mu_u - \mu_d)/2$, 
where $\mu_u$ and $\mu_d$ are the chemical potential for 
u and d quarks, respectively.
For the case with non-zero isospin and zero quark 
chemical potentials, $\mu_q= (\mu_u + \mu_d)/2$ =0, i.e.
$\mu_u = -\mu_d = \mu_I$, 
the quark determinant is real and positive because
\begin{eqnarray}
\det M(\mu_u) \det M(\mu_d)= \det M(\mu_I) \det M(-\mu_I)=
\det M(\mu_I) (\det M(\mu_I))^{*}= |\det M(\mu_I)|^2
\end{eqnarray}
where we used an identity at finite $\mu_q$, 
$\gamma_5 M(\mu_q) \gamma_5 = M(-\mu_q)^{\dagger}$. 
Therefore, Monte-Carlo simulations are possible for this case 
\cite{SS01}, and the simulations with the isospin chemical potential 
have been performed in Ref.~\cite{KS02,KS05,KS06,Ta04,NT03}. 
It may be important toward the understanding of QCD at finite density
to consider the difference between the phase diagram with 
non-zero baryon chemical and that with non-zero 
isospin chemical potential, 

The reweighting factor $\bar{R}$, 
i.e. the dashed line in Fig.~\ref{fig:rmu}, corresponds 
to the reweighting factor of the isospin chemical potential 
$R(P, \mu_I)$ for each $\mu_I/T$ because the quark determinant is 
$|\det M(\mu_q)|^2$. 
It is found from Fig.~\ref{fig:rmu} that  
the slope and the curvature of $\ln R$ around $P \sim 0.82$ 
for the isospin chemical potential are smaller than those for 
the quark chemical potential. 
This means that the value of $\mu_I/T$ where the second order 
phase transition appears by canceling the curvatures of 
$\ln w(P,\beta)$ and $\ln R(P, \mu_I)$ is larger than 
the critical point of $\mu_q/T$.
It is suggested in Ref.~\cite{KS06} that there is no first 
order phase transition region in the low density regime of 
QCD with non-zero $\mu_I/T$.
Although more quantitative estimations of the reweighting factor are 
needed to confirm the existence of the first order transition line, 
our argument may be related to their result.

Furthermore, in the case of the approximation up to $O(\mu_{q,I}^2)$, 
$R(P, \mu_q)$ and $R(P, \mu_I)$ have a close relation to 
the quark number susceptibility $\chi_q$ and 
isospin susceptibility $\chi_I$ at $\mu_{q,I}=0$. 
Using the equations (\ref{eq:absdet}), (\ref{eq:theta}) and (\ref{eq:a2}),
\begin{eqnarray}
\ln R(P, \mu_q) & \approx & \ln \left\langle 
\exp \left\{ \frac{1}{2} N_{\rm f} {\rm Re} 
\frac{\partial^2 (\ln \det M)}{\partial (\mu_q/T)^2} 
\left( \frac{\mu_q}{T} \right)^2 \right\} \right\rangle_P
-\frac{1}{2} \left\langle \left( N_{\rm f} {\rm Im} 
\frac{\partial (\ln \det M)}{\partial (\mu_q/T)} 
\frac{\mu_q}{T} \right)^2 \right\rangle_P \nonumber \\
& \approx & \frac{1}{2} \left[ \left\langle 
N_{\rm f} \frac{\partial^2 (\ln \det M)}{\partial (\mu_q/T)^2} 
+\left( N_{\rm f} \frac{\partial (\ln \det M)}{\partial (\mu_q/T)} 
\right)^2 \right\rangle_P 
\right] \left( \frac{\mu_q}{T} \right)^2 
\end{eqnarray}
in this approximation, 
and when the effect from $\theta$ is omitted, we find 
\begin{eqnarray}
\ln R(P, \mu_I) = \ln \bar{R}(P, \mu_q) \approx \frac{1}{2} 
\left\langle N_{\rm f} 
\frac{\partial^2 (\ln \det M)}{\partial (\mu_q/T)^2} 
\right\rangle_P \left( \frac{\mu_q}{T} \right)^2, 
\end{eqnarray}
where $\langle \cdots \rangle_{P'} = \langle \cdots \delta(P'-P) \rangle 
/ \langle \delta(P'-P) \rangle$. 
These are related to $\chi_q /T^2$ and $\chi_I /T^2$ as functions of 
$\beta$ (temperature) by the following equations 
\begin{eqnarray}
\frac{\chi_q}{T^2} (T, \mu_{q,I}=0) &=& 
\frac{N_t^3}{N_s^3} \frac{1}{\cal Z} \int 
\left\langle N_{\rm f} 
\frac{\partial^2 (\ln \det M)}{\partial (\mu_q/T)^2} 
+\left( N_{\rm f} \frac{\partial (\ln \det M)}{\partial (\mu_q/T)} 
\right)^2 \right\rangle_P w(P,\beta) \ dP, \\
\frac{\chi_I}{T^2} (T, \mu_{q,I}=0) &=& 
\frac{N_t^3}{N_s^3} \frac{1}{\cal Z} \int 
\left\langle N_{\rm f} \frac{\partial^2 (\ln \det M)}{\partial (\mu_q/T)^2} 
\right\rangle_P w(P,\beta) \ dP.
\end{eqnarray}
From these equations, the similarity between the figures for 
$R(P, \mu_{q,I})$ and those of 
the quark number and isospin susceptibilities 
can be easily understood in the regime where the Taylor expansion 
is valid. 
As shown in Fig.~\ref{fig:hist}, $w(P,\beta)$ is a Gaussian function 
having a sharp peak. Therefore, Fig.~\ref{fig:rmu} is quite 
similar to Fig.~1 in Ref.~\cite{BS05} if we replace the horizontal 
axis $P$ by $T/T_c (\beta)$.
As we have discussed, the positive curvature in the 
$P$ dependence of $\ln R(P, \mu_q)$ is required 
for the appearance of the critical endpoint. 
It is found that the positive curvature is related closely to 
the rapid increase of the quark number susceptibility near 
the pseudo-critical temperature at $\mu_q=0$.

Here, it should be noted that $\chi_I /T^2$ is always larger than 
$\chi_q /T^2$ at $\mu_q=0$ because 
$\partial (\ln \det M)/ \partial (\mu_q/T)$ is purely imaginary, 
and thus $(\partial (\ln \det M)/ \partial (\mu_q/T))^2$ is negative. 
Moreover, both susceptibilities approach the same value 
in the high temperature limit. 
Hence, the variation of $\chi_I /T^2$ around the transition point 
would be milder than that of $\chi_q /T^2$, corresponding 
to the behavior of $R(P, \mu_q)$ and $R(P, \mu_I)$. 
This may explain the difference between the phase diagrams with 
finite $\mu_q$ and finite $\mu_I$. 
Furthermore, in the framework of the hadron resonance gas model at 
low temperature, the isospin susceptibility correspond to 
fluctuations of pions, and the pion mass is more sensitive to 
the quark mass than baryon masses. Therefore, when the quark mass 
is decreased, the pion mass becomes smaller and the fluctuation 
becomes larger at low temperature. 
This suggests the change of $\chi_I /T^2$ 
around $T_c$ may be milder at small quark mass, i.e. 
the difference between $\ln R(P, \mu_q)$ and $\ln R(P, \mu_I)$ 
becomes large at small quark mass.

For more precise arguments on the phase structure, more accurate 
evaluations of $R(P, \mu_q)$ and $R(P, \mu_I)$ are required 
increasing the number of terms in the Taylor expansion of $\ln \det M$. 
However, the qualitative property that
the critical value of $\mu_q/T$ in the $(T, \mu_q)$ plane is smaller 
than the critical $\mu_I/T$ in the $(T, \mu_I)$ plane 
can be understood by the well-known properties of the quark (baryon) 
number and isospin susceptibilities combined with 
the argument of the effective potential.

\section{Conclusions}
\label{sec:conc}

We have discussed the phase structure of lattice QCD at non-zero density. 
The probability distribution as a function of the plaquette 
was estimated at non-zero temperature and chemical potential using 
the data obtained with two-flavors of p4-improved staggered quarks 
in Ref.~\cite{BS05}. 
In this analysis, we have adopted two approximations. 
One is that we estimate $\ln \det M$ from the data of a Taylor expansion 
up to $O(\mu_q^6)$. 
Terms of higher order than $\mu_q^6$ are omitted.
We have estimated the range where this approximation is valid 
and studied in the reliability range.
The second approximation is an assumption on the probability 
distribution for the complex phase. We have assumed the 
distribution function to be a Gaussian function. 
This assumption will be valid for sufficiently large volume and 
we have checked that the distribution is well-approximated by 
a Gaussian function for the data used in this analysis. 

In spite of the use of these approximations, it is found that 
the shape of the effective potential which is of Gaussian type at 
$\mu_q=0$ changes to a double-well type at large $\mu_q/T$. 
This property is related closely to a well-known behavior of the quark 
number susceptibility at $\mu_q=0$, i.e. the rapid increase near 
the phase transition point. 
For the quantitative estimation of the endpoint of the first order 
phase transition, further investigation must be needed. 
However, this argument strongly suggests the existence of 
the first order phase transition line in the $(T, \mu_q)$ plane. 

We also discussed the difference between QCD with a quark chemical 
potential and QCD with an isospin chemical potential, and found that 
the critical value of the quark chemical potential seems to be smaller 
than that of the isospin chemical potential.

\section*{Acknowledgments}
I would like to thank 
F. Karsch, K. Kanaya, T. Hatsuda, S. Aoki, T. Izubuchi and K. Fukushima 
for discussions and comments.
This work has been authored under contract DE-AC02-98CH10886
with the U.S. Department of Energy.
I also wish to thank the Sumitomo Foundation for their 
financial assistance (No.~050408) and the Yukawa Institute 
for Theoretical Physics at Kyoto University for discussions 
during the YITP workshops YITP-W-06-07 and YKIS2006.

%, Grants-in-Aid of the Japanese MEXT (No.~18740134)


\begin{thebibliography}{99}
\bibitem{FK1}
Z. Fodor and S. Katz, 
Phys. Lett. B {\bf 534}, 87 (2002).
%%CITATION = HEP-LAT 0104001;%%

\bibitem{FK2}
Z. Fodor and S. Katz, 
JHEP {\bf 0203}, 014 (2002);
JHEP {\bf 0404}, 050 (2004).
%%CITATION = HEP-LAT 0106002;%%
%%CITATION = HEP-LAT 0402006;%%

\bibitem{BS02}
C.R. Allton, S. Ejiri, S.J. Hands, O. Kaczmarek, F. Karsch, E. Laermann, 
Ch. Schmidt and L. Scorzato, Phys. Rev. D {\bf 66}, 074507 (2002).
%%CITATION = HEP-LAT 0204010;%%

\bibitem{dFP1}
P. de Forcrand and O. Philipsen, Nucl. Phys. B {\bf 642}, 290 (2002).
%%CITATION = HEP-LAT 0205016;%%

\bibitem{dEL1}
M. D'Elia and M.-P. Lombardo, Phys. Rev. D {\bf 67}, 014505 (2003).
%%CITATION = HEP-LAT 0209146;%%

\bibitem{BS03}
C.R. Allton, S. Ejiri, S.J. Hands, O. Kaczmarek, F. Karsch, E. Laermann 
and C. Schmidt, Phys. Rev. D {\bf 68}, 014507 (2003).
%%CITATION = HEP-LAT 0305007;%%

\bibitem{BS05}
C.R. Allton, M. D\"{o}ring, S. Ejiri, S.J. Hands, O. Kaczmarek, F. Karsch, 
E. Laermann and K. Redlich, Phys. Rev. D {\bf 71}, 054508 (2005).
%%CITATION = HEP-LAT 0501030;%%

\bibitem{isen06}
S. Ejiri, F. Karsch, E. Laermann and C. Schmidt, 
Phys. Rev. D {\bf 73}, 054506 (2006).

\bibitem{MI06}
C.Bernard {\it et. al.}, {\tt hep-lat/0610017}.

\bibitem{AY}
M. Asakawa and K. Yazaki, Nucl. Phys. A {\bf 504}, 668 (1989).

\bibitem{Bard}
A. Barducci, R. Casalbuoni, S. De Curtis, R. Gatto and G. Pettini,
Phys. Lett. B {\bf 231}, 463 (1989); Phys. Rev. D {\bf 41}, 1610 (1990);
A. Barducci, R. Casalbuoni, G. Pettini and R. Gatto,
Phys. Rev. D {\bf 49}, 426 (1994).

\bibitem{SRS}
M. Stephanov, K. Rajagopal and E. Shuryak, 
Phys. Rev. Lett. {\bf 81}, 4816 (1998).

\bibitem{ej06}
S. Ejiri, Phys. Rev. D {\bf 73}, 054502 (2006).
%%CITATION = HEP-LAT 0506023;%%

\bibitem{GG2}
R.V. Gavai and S. Gupta, Phys. Rev. D {\bf 71} 114014 (2005).

\bibitem{crtpt}
Ch. Schmidt, C.R. Allton, S. Ejiri, S.J. Hands, O. Kaczmarek, F. Karsch and E.
Laermann, Nucl. Phys. B (Proc. Suppl.) {\bf 119}, 517 (2003);
F. Karsch, C.R. Allton, S. Ejiri, S.J. Hands, O. Kaczmarek, E. Laermann 
and Ch. Schmidt, Nucl. Phys. B (Proc. Suppl.) {\bf 129}, 614 (2004);
S. Ejiri, C.R. Allton, S.J. Hands, O. Kaczmarek, F. Karsch, E. Laermann 
and Ch. Schmidt, Prog. Theor. Phys. Suppl. {\bf 153}, 118 (2004).
%{\tt hep-lat/0309116}.
%{\tt hep-lat/0209009}.
%%CITATION = HEP-LAT 0209009;%%

\bibitem{dFP2} 
P. de Forcrand and O. Philipsen, Nucl. Phys. B {\bf 673}, 170 (2003);
JHEP {\bf 0701}, 077 (2007)
%%CITATION = HEP-LAT 0307020;%%

\bibitem{dEL2}
M. D'Elia and M.-P. Lombardo, Phys. Rev. D {\bf 70}, 074509 (2004).

\bibitem{KS05}
J.B. Kogut and D.K. Sinclair, {\tt hep-lat/0509095}.

\bibitem{KS06}
J.B. Kogut and D.K. Sinclair, {\tt hep-lat/0609041}. 

\bibitem{FKS07}
Z. Fodor, S. Katz and C. Schmidt,
JHEP {\bf 0703}, 121 (2007).

\bibitem{Swen88} A.M. Ferrenberg and R.H. Swendsen, 
Phys.\ Rev.\ Lett.\ {\bf 61}, 2635 (1988);
Phys.\ Rev.\ Lett.\ {\bf 63}, 1195 (1989).

\bibitem{Bar}
I.M. Barbour, S.E. Morrison, E.G. Klepfish, J.B. Kogut, M.-P. Lombardo, 
Phys. Rev. D {\bf 56}, 7063 (1997).

\bibitem{spli}
K. Splittorff, PoS {\bf LAT2006}, 023 (2006); 
K. Splittorff and J.J.M. Verbaarschot, 
Phys. Rev. Lett. {\bf 98}, 031601 (2007).

\bibitem{ej04}
S. Ejiri, Phys. Rev. D {\bf 69}, 094506 (2004). 

\bibitem{KLS01}
F. Karsch, E. Laermann and C. Schmidt, 
Phys. Lett. B {\bf 520} 41 (2001).

\bibitem{Gock88} 
A. Gocksch, Phys. Rev. Lett. {\bf 61}, 2054 (1988).

\bibitem{JN01}
K.N. Anagnostopoulos and J. Nishimura,
Phys. Rev.  D {\bf 66}, 106008 (2002);
%  [arXiv:hep-th/0108041]
J. Ambjorn, K.N. Anagnostopoulos, J. Nishimura and J.J.M. Verbaarschot,
JHEP {\bf 0210}, 062 (2002).
%  [arXiv:hep-lat/0208025].

\bibitem{Ta04}
T. Takaishi,
Mod. Phys. Lett. A {\bf 19},909 (2004)

\bibitem{KRT}
F. Karsch, K. Redlich and A. Tawfik,
Eur. Phys. J. C {\bf 29} (2003) 549;
%%CITATION = HEP-PH 0303108;%%
Phys. Lett. B {\bf 571} (2003) 67.
%%CITATION = HEP-PH 0306208;%%

\bibitem{SS01} 
D.T. Son and M.A. Stephanov, Phys. Rev. Lett. {\bf 86}, 592 (2001).

\bibitem{KS02}
J.B. Kogut and D.K. Sinclair, Phys. Rev. D {\bf 66}, 034505 (2002).

\bibitem{NT03}
A. Nakamura and T. Takaishi,
Nucl. Phys. (Proc. Suppl.) {\bf 129}, 629 (2004).

\end{thebibliography}
\end{document}